\documentclass[11pt]{article}
\usepackage{blois,epsfig}

\def\Journal#1#2#3#4{{#1} {\bf #2}, #3 (#4)}

\def\PLB{{\em Phys. Lett.}  B}
\def\PRL{\em Phys. Rev. Lett.}

\def\prep{{\em Phys. Rep.}~}

\def\apj{{\em Astrophys. J.}~} 
\def\apjs{{\em Astrophys. J. Suppl.}~}  
  
\def\nat{{\em Nature}~}

\def\eg{{e.g.},~}  
\def\etal{{\em et al.} }  
\def\4he{$^4$He}  
\def\3he{$^3$He}  
\def\7li{$^7$Li}  
\def\Yp{Y$_{\rm P}$}  
\def\yd{$y_{\rm D}$~}  
\def\y3{$y_{3}$~}  
\def\hii{H\thinspace{$\scriptstyle{\rm II}$}~}  
\def\hi{H\thinspace{$\scriptstyle{\rm I}$}~}  
\def\di{D\thinspace{$\scriptstyle{\rm I}$}~}  
\newcommand\la{\lower0.6ex\vbox{\hbox{\ensuremath{\buildrel{\textstyle<}\over{\sim}\  
}}}}  
\newcommand\ga{\lower0.6ex\vbox{\hbox{\ensuremath{\buildrel{\textstyle>}\over{\sim}\  
}}}}  
  
\newcommand{\obh}{\ensuremath{\Omega_{\rm B} h^2\;}}

\newcommand{\omb}{\ensuremath{\Omega_{\rm B}\;}}

\newcommand{\be}{\begin{equation}}  
\newcommand{\ee}{\end{equation}}  
\newcommand{\Deln}{\ensuremath{\Delta N_\nu\;}}  
\def\Nnu{$N_{\nu}$~}  
\newcommand{\epm}{\ensuremath{e^{\pm}\;}}  
  
\def\be{\begin{equation}}
\def\ee{\end{equation}}
\def\bea{\begin{eqnarray}}
\def\eea{\end{eqnarray}}

\begin{document}
\vspace*{4cm}
\title{THE BARYON BUDGET FROM BBN AND THE CBR}

\author{ GARY STEIGMAN }

\address{Department of Physics, The Ohio State University,\\
174 West 18th Avenue\\
Columbus, OH 43210, USA}

\maketitle\abstracts{
A  key pillar of modern cosmology, Big Bang Nucleosynthesis (BBN) offers 
a probe of the particle content and expansion rate of the Universe a mere 
few minutes after the beginning.  When compared with the BBN predictions,
the observationally inferred primordial abundances of deuterium and helium-4 
provide an excellent baryometer and chronometer respectively.  Several 
hundred thousand years later, when the Cosmic Background Radiation (CBR) 
photons began progagating freely, the spectrum of temperature fluctuations 
imprinted on them also encoded information about the baryon density and the 
expansion rate.  Comparing the constraints imposed by BBN with those from 
the CBR reveals a consistent picture of the Universe at these two very 
widely separated epochs.  Combining these probes leads to new, tighter 
constraints on the baryon density at present and on possible new physics 
beyond the standard model of particle physics.
}

\section{Introduction}\label{sec:intro}

As the hot, dense, early Universe rushed to expand and cool, it briefly
passed through the epoch of big bang nucleosynthesis (BBN), leaving behind
as relics the first complex nuclei: deuterium, helium-3, helium-4, and
lithium-7.  The abundances of these relic nuclides were determined by the 
competition between the relative densities of nucleons (baryons) and photons 
and also by the universal expansion rate.  In particular, while deuterium 
is an excellent baryometer, \4he provides an accurate chronometer.  Nearly 
400 thousand years later, when the cosmic background radiation (CBR) had 
cooled sufficiently to allow neutral atoms to form, releasing the CBR from 
the embrace of the ionized plasma of protons and electrons, the spectrum 
of temperature fluctuations imprinted on the CBR preserved the values of 
the contemporary baryon and radiation densities, along with the universal 
expansion rate at that epoch.  As a result, the relic abundances of the 
light nuclides and the CBR temperature fluctuation spectrum provide 
invaluable windows on the early evolution of the Universe as well as 
sensitive probes of its particle content.

The fruitful interplay between theory and data has been key to the enormous
progress in cosmology in recent times.  As new, more precise data became
available, models have had to be rejected or refined.  It is anticipated
this this process will -- indeed, should -- continue.  Therefore, this
review of the baryon content of the Universe as revealed by BBN and the 
CBR is but a signpost on the road to a more complete understanding of 
the history and evolution of the Universe (for a related review, with 
more detailed discussion and  a more extensive bibliography, see 
Steigman~\cite{steigman03}).  By highlighting the current successes of 
the present ``standard'' model along with some of the challenges to it, 
I hope to identify those areas of theoretical and observational work 
which will contribute to progress towards the goal of understanding the 
Universe, its past, present, and future.

\section{BBN And The Baryon Density}\label{sec:bbn}

The hot, dense, early Universe is a hostile environment for complex nuclei. 
At sufficiently high temperatures ($T ~\ga 80$~keV), when all the nucleons 
(baryons) were either neutrons or protons, their relative abundance was 
regulated by the weak interaction; the higher mass of the neutron ensures 
that protons dominate (in the absence of a chemical potential for the 
electron-type neutrinos $\nu_{e}$). Below $\sim$~80~keV, the Universe has 
cooled sufficiently that the cosmic nuclear reactor can begin in earnest, 
building the lightest nuclides D, \3he, \4he, and \7li. Very rapidly, D and 
\3he (also $^3$H) are burned to \4he, the most tightly bound of the light 
nuclides.  The absence of a stable mass-5 nuclide ensures that in the 
expanding, cooling, early Universe, the abundances of heavier nuclides are 
greatly suppressed.  By the time the temperature has dropped below $\sim 
30$~keV, a time comparable to the neutron lifetime, the average thermal 
energy of the nuclides and nucleons is too small to overcome the coulomb 
barriers, any remaining free neutrons decay, and BBN ceases.    

\begin{figure}
\centering
 \epsfysize=4.0truein
  \epsfbox{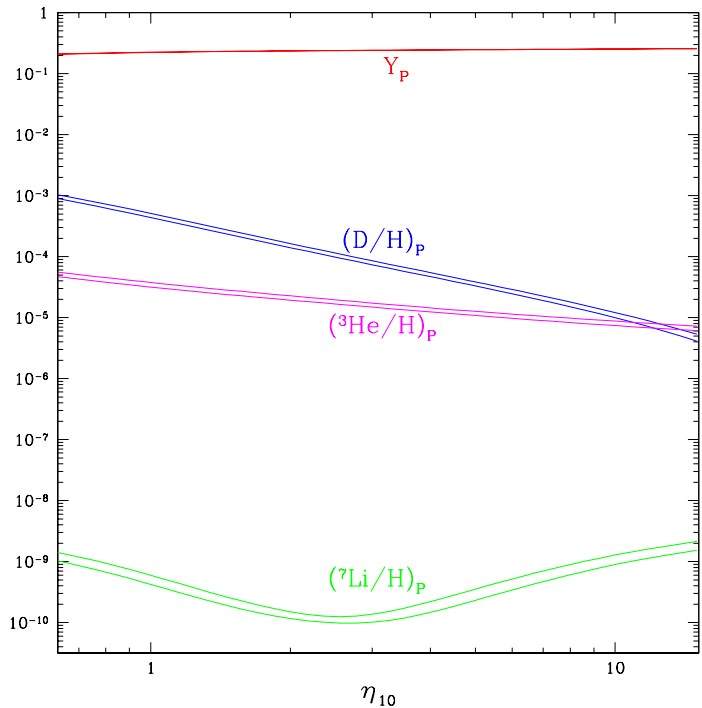}
\caption{The SBBN-predicted primordial abundances of D, \3he, and 
\7li by number with respect to hydrogen, and the \4he mass fraction 
Y$_{\rm P}$, as a function of the nucleon (baryon) abundance parameter 
$\eta_{10}$.  The bands reflect the theoretical uncertainties 
($\pm 1\sigma$) in the BBN predictions.
\label{fig:schrplot}}
\end{figure}

In the expanding Universe the number densities of all particles decrease 
with time, so that the magnitude of the baryon density (or that of any other 
particle) has no meaning without also specifying {\it when it is measured}.  
To quantify the universal abundance of baryons, it is best to compare the
baryon number density $n_{\rm B}$ to the CBR photon number density 
$n_{\gamma}$.  After \epm pairs have annihilated the ratio of these number 
densities remains effectively constant as the Universe evolves.  This ratio 
$\eta \equiv n_{\rm B}/n_{\gamma}$ is very small, so that it is convenient 
to define a quantity of order unity,
\be
\eta_{10} \equiv 10^{10}(n_{\rm B}/n_{\gamma}) = 274\,\Omega_{\rm B}h^{2},
\label{eq:eta10}
\ee
where \omb is the ratio (at present) of the baryon density to the critical
density and $h$ is the present value of the Hubble parameter in units of
100 km s$^{-1}~$Mpc$^{-1}$.  The relic abundances of D, \3he, and \7li 
shown in Figure~\ref{fig:schrplot} are {\it rate limited}, determined by 
the competition between the early Universe expansion rate and the nucleon 
density.  Any of these three nuclides is a potential baryometer.  

In contrast to the synthesis of the other light nuclides, once BBN 
commences the reactions building \4he are so rapid that its relic 
abundance is not rate limited but, rather, is limited by the availability 
of neutrons.  To a very good approximation, the relic abundance of \4he 
is set by the neutron abundance at the beginning of BBN.  The neutron 
abundance (relative to protons) is determined by the competition between 
the universal expansion rate (the Hubble parameter, $H$) and the weak 
interaction rates ($p + e^{-} \longleftrightarrow n + \nu_{e}, ~n + e^{+} 
\longleftrightarrow p + \bar{\nu}_{e}, ~n \longleftrightarrow p + e^{-} 
+ \bar{\nu}_{e}\,$), which are well constrained (in the absence of large 
$\nu_{e}$ chemical potentials) by the neutron lifetime ($\tau_{n} = 885.7 
\pm 0.8$~s).  As a result, the primordial mass fraction of \4he, Y$_{\rm P}$, 
while a relatively insensitive baryometer (see Figure~\ref{fig:schrplot}), 
is an excellent, early-Universe chronometer.  

The expansion rate (measured by the Hubble parameter $H$) is related to 
the energy density through the Friedman equation ($H^{2}={8\pi G \over 
3}\rho$).  During BBN the Universe is ``radiation-dominated'' (RD); 
the energy density ($\rho_{\rm R}$) is dominated by the relativistic 
particles present. For the standard model (SBBN) there are three families 
of light (relativistic) neutrinos (N$_{\nu} = 3$).  To explore models 
with nonstandard, early (RD) expansion rates it is convenient to introduce 
the {\it expansion rate factor} $S \equiv H'/H = t/t'$.  One possible 
origin for a nonstandard expansion rate ($S \neq 1$) is a modification
of the energy density by the presence of ``extra'' relativistic particles 
$X$: $\rho_{\rm R} \rightarrow \rho_{\rm R}' = \rho_{\rm R} + \rho_{X}$.  
If the additional energy density is normalized to that which would be 
contributed by additional flavors of (decoupled) neutrinos (Steigman, 
Schramm, \& Gunn~\cite{ssg}), $\rho_{X} \equiv \Delta N_{\nu}\rho_{\nu}$,
then $N_{\nu} = 3 + \Delta N_{\nu}$. It must be emphasized that the 
fundamental physical parameter is $S$, the expansion rate factor, 
which may be related to \Deln ({\it nonlinearly}) by
\be
S_{pre} \equiv (H'/H)_{pre} = (1 + 0.163\Delta N_{\nu})^{1/2}\,; \, \, \, 
S_{post} \equiv (H'/H)_{post} = (1 + 0.135\Delta N_{\nu})^{1/2},
\label{eq:sx}
\ee
where the subscripts ``$pre$'' and ``$post$'' reflect the values prior to,
and after \epm annihilation respectively.  However, any term in a modified 
(nonstandard) Friedman equation which scales like radiation (decreases as 
the fourth power of the scale factor), such as may be due to higher 
dimensional effects as in the Randall-Sundrum~\cite{rs} model, will change 
the standard-model expansion rate ($S \neq 1$) and may be related to an 
{\it equivalent} \Deln (which may be {\it negative} as well as positive; 
both $S > 1$ and $S < 1$ are possible) through Eq.~\ref{eq:sx}.

The qualitative effects of a nonstandard expansion rate on the relic
abundances of the light nuclides may be understood with reference to
Figure~\ref{fig:schrplot}.  For the baryon abundance range of interest 
($1 ~\la \eta_{10} ~\la 10$) the relic abundances of D and \3he are
decreasing functions of $\eta$, revealing that in this range D and \3he
are being destroyed.  A faster than standard expansion ($S > 1$) leaves
less time for this destruction so that more D and \3he would survive.
The same is true for \7li for low values of $\eta$, where the abundance
of \7li is a decreasing function of $\eta$.  However, at higher values
of $\eta$, where the \7li abundance increases with $\eta$, less time
available results in less production and a {\it smaller} \7li relic
abundance.  Except for dramatic changes to the early-Universe expansion
rate, these effects on the relic abundances of D, \3he, and \7li are 
subdominant to their dependences on the baryon density.  Not so for 
\4he, whose relic abundance is very weakly (logarithmically) dependent 
on the baryon density, but very strongly dependent on the early-Universe 
expansion rate.  A faster expansion leaves more neutrons available to be 
incorporated into \4he; to a good approximation, $\Delta$Y $\approx 
0.16\,(S-1)$.  It is clear then that if \4he is paired with any of 
the other light nuclides, together they can constrain both the baryon 
density (\obh or $\eta$) and the early-Universe expansion rate ($S$ 
or $\Delta$N$_{\nu}$).

\section{The Best BBN Baryometer?}\label{sec:d}

At present, deuterium is the baryometer of choice.  As the Universe evolves,
structure forms, and gas is cycled through stars.  Because of its very low
binding energy, deuterium is completely destroyed in any gas which passes 
through stars; in the post-BBN Universe the D abundance never exceeds that
emerging from BBN.  Because the evolution of D is simple, monotonically 
decreasing since BBN, observations of D anywhere, anytime, yield a lower 
bound to its primordial abundance.  Observations of D in ``young'' systems 
(high redshift or low metallicity), should reveal a deuterium ``plateau'' 
at its primordial abundance.  In addition to its simple post-BBN evolution, 
deuterium is the baryometer of choice also because its primordial abundance 
($y_{\rm D} \equiv 10^{5}(D/H)_{\rm P}$) is sensitive to the baryon density, 
$y_{\rm D} \propto \eta^{-1.6}$; as a result, a $\sim 10\%$ measurement 
of \yd will lead to a $\sim 6\%$ determination of $\eta$ (or $\Omega_
{\rm B}h^{2}$).

In contrast to D, the post-BBN evolution of \3he and \7li are complex.  
For example, \3he is destroyed in the hotter interiors of all but the 
least massive stars, but is preserved in the cooler, outer layers of 
most stars.  In addition, hydrogen burning in low mass stars results 
in the production of significant amounts of {\it new} \3he \cite{3he}. 
To follow the post-BBN evolution of \3he, it is necessary to account 
for all these effects -- quantitatively -- in the material returned by 
stars to the interstellar medium (ISM).  As indicated by the existing 
Galactic data~\cite{gg}, a very delicate balance exists between net 
production and net destruction of \3he in the course of the evolution 
of the Galaxy.  Thus, aside from noting the excellent qualitative 
agreement between the SBBN prediction and the observed \3he abundances, 
\3he has little role to play, at present, as a quantitatively useful 
baryometer. A similar scenario may be sketched for \7li.  As a weakly 
bound nuclide, it is easily destroyed when cycled through stars.  The 
high lithium abundances observed in the ``super-lithium-rich red giants'' 
provide direct evidence that at least some stars do produce post-BBN 
lithium, but a key unsolved issue is how much of this newly-synthesized 
lithium is actually returned to the ISM.  Furthermore, the quest for 
observations of nearly primordial lithium is limited to the oldest, 
most metal-poor stars in the Galaxy, stars that have had the most time 
to redistribute -- and destroy or dilute -- their surface lithium 
abundances.  At present it seems that the most fruitful approach is 
to use the BBN-predicted lithium abundance, compared to that inferred 
from observations of the oldest stars, to learn about stellar evolution 
rather than to use stellar observations to constrain the BBN-inferred 
baryon density.

\subsection{Deuterium}\label{subsec:deut}

\begin{figure}
\centering
 \epsfysize=4.0truein
  \epsfbox{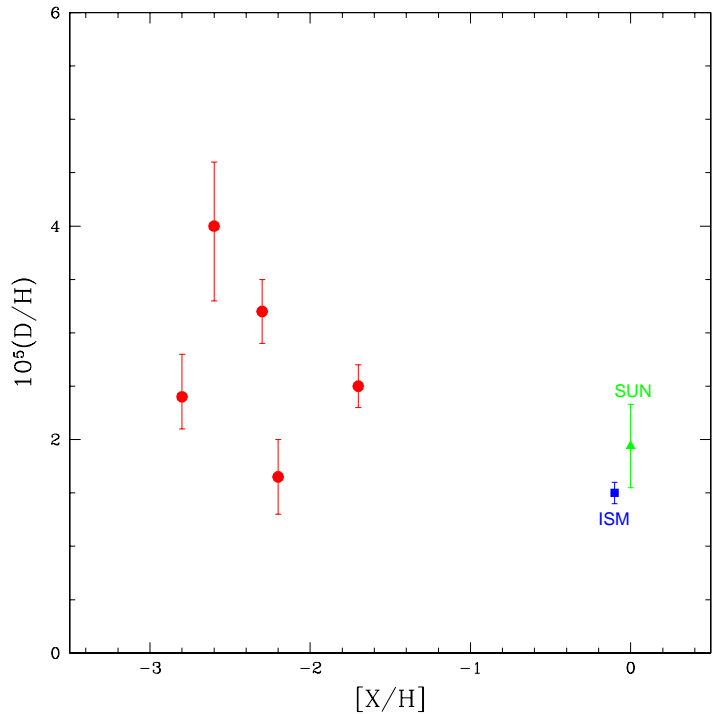}
\caption{The deuterium abundance by number with respect to hydrogen, 
versus the metallicity (relative to solar on a log scale), from 
observations (as of early 2003) of QSOALS (filled circles).  ``X'' 
is usually Si.  Also shown for comparison are the D abundances 
for the local ISM (filled square) and the solar system (``Sun''; 
filled triangle).    
\label{fig:dvssi}}  
\end{figure}

In pursuit of the most nearly primordial D abundance, it is best to
concentrate on ``young'' systems, in the sense of those which
have experienced the least evolution.  Thus, while observations of D
in the solar system and/or the local ISM provide useful {\it lower}
bounds to the primordial D abundance, it is observations of deuterium
(and hydrogen) absorption in a handful (so far) of high redshift, low
metallicity, QSO absorption-line systems (QSOALS) which provide the most 
relevant data.  However, inferring the primordial D abundance from the 
QSOALS has not been straightforward, with some abundance claims having 
been withdrawn or revised.  Presently there are only five QSOALS with 
reasonably firm deuterium detections~\cite{deut}; the derived abundances 
of D are shown in Figure~\ref{fig:dvssi} along with the corresponding 
solar system and ISM D abundances.

As is clear from Figure~\ref{fig:dvssi}, there is significant dispersion 
among the derived D-abundances at low metallicity.  The data fail to reveal 
the anticipated deuterium plateau, suggesting that systematic errors may 
be present, contaminating some of the determinations of the \di and/or \hi 
column densities.  Since the \di and \hi absorption spectra are identical, 
except for the wavelength/velocity offset resulting from the heavier 
reduced mass of the deuterium atom, an \hi ``interloper'', a low-column 
density cloud shifted by $\sim 81$~km s$^{-1}$ with respect to the main 
absorbing cloud, would masquerade as D\thinspace{$\scriptstyle{\rm I}$}.  
If this is not accounted for, a D/H ratio which is too high would be 
inferred.  Since there are more low-column density absorbers than those 
with high \hi column densities, absorption-line systems with somewhat 
lower \hi column density (\eg Lyman-limit systems) are more susceptible 
to this contamination than are the higher \hi column density absorbers 
(\eg damped Ly$\alpha$ absorbers).  However, for the damped Ly$\alpha$ 
absorbers, an accurate determination of the \hi column density requires 
an accurate placement of the continuum, which could be compromised by 
interlopers.  This might lead to an overestimate of the \hi column density 
and a concomitant underestimate of D/H (J. Linsky, private communication).  
There is the possibility that each of these effects, separately or in 
combination,  may have contaminated some of the current data, leading to 
the excessive dispersion seen in Figure~\ref{fig:dvssi} (provided there 
is, indeed, a true deuterium ``plateau'' at low metallicity).  To utilize 
the current data I follow the lead of O'Meara \etal \cite{deut} and Kirkman 
\etal \cite{deut} and adopt for the primordial D abundance the weighted 
mean of the D abundances for the five lines of sight~\cite{deut}, while 
the dispersion in the data is used to set the error in $y_{\rm D}$: \yd = 
$2.6 \pm 0.4$.  It is worth noting that for the same data Kirkman \etal 
\cite{deut} derive a slightly higher mean D abundance: \yd = 2.74.  The 
difference is due to their first finding the mean of log(y$_{\rm D}$) 
and then using it to compute the mean D abundance (\yd $ \equiv 
10^{\langle \log(y_{\rm D})\rangle}$). 
   
\section{Helium-4: The Best BBN Chronometer}\label{sec:he4}

While D, \3he, and \7li are all potential BBN baryometers, the relic
abundance of \4he is too weakly dependent on the universal density of
baryons to permit it to be used as a effective baryometer.  However,
the primordial abundance of \4he is sensitive to the early-Universe
expansion rate.  The primordial \4he mass fraction, \Yp, is the BBN 
chronometer of choice.  The net effect of post-BBN evolution was to 
burn hydrogen to helium, increasing the \4he mass fraction above its 
primordial value (Y $>$ \Yp).  As with D, although \4he is constrained 
by solar system and ISM (Galactic \hii regions) data, the key to the 
\Yp ~determinations are the observations of \4he in nearly unevolved, 
metal-poor regions.  To date they are limited to the observations of 
helium and hydrogen recombination lines in extragalactic \hii regions 
\cite{hii}.  Unfortunately, at present the \hii region \4he data has 
an even greater dispersion in the derived \Yp ~values than that for 
primordial D, ranging from \Yp~= 0.234$\pm 0.003$ to \Yp~= 0.244$\pm 
0.002$, suggesting that current estimates are not limited by statistics, 
but by uncorrected systematic errors~\cite{vgs}.  Here, I adopt the 
compromise proposed in Olive, Steigman, and Walker~\cite{osw}: 
\Yp~= 0.238$\pm 0.005$; for further discussion and references, 
see Steigman~\cite{steigman03}.

\section{Testing BBN}\label{test}

\begin{figure}
\centering
 \epsfysize=4.0truein
  \epsfbox{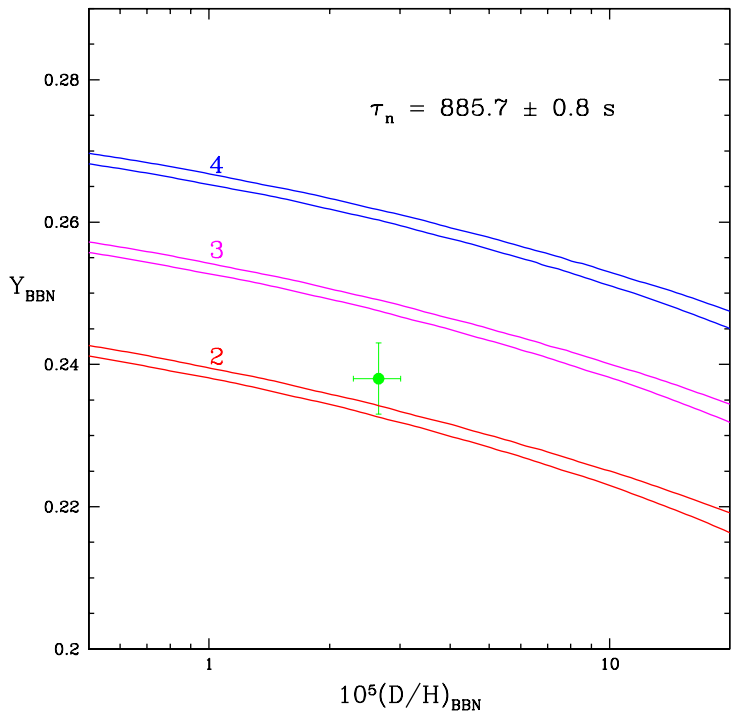}
\caption{The BBN-predicted \4he mass fraction Y$_{\rm BBN}$ versus 
the corresponding BBN-predicted deuterium abundance \yd $\equiv 
10^{5}($D/H)$_{\rm BBN}$ for \Nnu = 2, 3, 4 (corresponding to $S$ 
= 0.915, 1, 1.078). The data point (filled circle with error bars) 
is for the D and \4he abundances adopted here (see the text).
\label{fig:hevsd234}}   
\end{figure} 

For SBBN, the baryon density corresponding to the D abundance adopted 
here ($y_{\rm D} = 2.6 \pm 0.4$) is $\eta_{10} = 6.1 ^{+0.7}_{-0.5}$, 
corresponding to \obh = $0.022^{+0.003}_{-0.002}$.  This is in outstanding 
agreement with the estimate of Spergel \etal \cite{wmap}, based largely 
on the new CBR ({\it WMAP}) data (Bennett \etal \cite{wmap}): \obh $ = 
0.0224 \pm 0.0009$.  For the baryon density determined by D, the 
SBBN-predicted abundance of \3he is $y_{3} = 1.0 \pm 0.1$, which is 
to be compared to the outer-Galaxy abundance of $y_{3} = 1.1 \pm 0.1$, 
suggested by Bania, Rood, \& Balser \cite{gg} to be nearly primordial.  
Again, the agreement is excellent; the concordance between SBBN and the 
CBR is spectacular.  

\begin{figure}[t]  
\centering
 \epsfysize=3.0truein
  \epsfbox{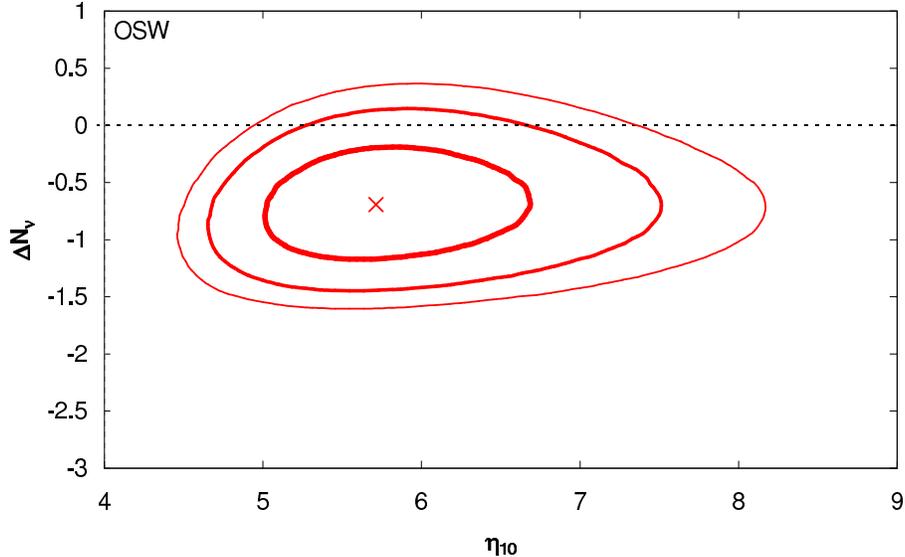}
\caption{The 1-, 2-, and 3-$\sigma$ contours in 
the $\eta$ -- \Deln plane for BBN and the adopted 
D and \4he abundances. 
\label{fig:bbncontours}}   
\end{figure}    

Tension between the data and SBBN arises with \4he.  In Figure~\ref
{fig:hevsd234} are shown the BBN-predicted \Yp ~versus \yd relations 
for SBBN (\Nnu = 3) as well as for two variations on SBBN (\Nnu = 2 
and 4).  Also shown in Figure~\ref{fig:hevsd234} are the primordial 
abundances of D and \Yp ~adopted here; see Sec.~\ref{subsec:deut} and 
Sec.~\ref{sec:he4}.  Given the very slow variation of \Yp ~with $\eta$, 
along with the very high accuracy of the SBBN-abundance prediction, 
the primordial abundance is tightly constrained: Y$_{\rm P}^{\rm SBBN} 
= 0.248 \pm 0.001$.  Agreement with the adopted value of Y$_{\rm P}^{\rm 
OSW} = 0.238 \pm 0.005$ is only at the $\sim 5\%$ level. This apparent 
challenge to SBBN is also an opportunity.  

As already noted, while the \4he abundance is insensitive to the baryon 
density, it is very sensitive to a nonstandard early expansion rate. 
Given the {\it pair} of primordial abundances \{\Yp, $y_{\rm D}$\} 
inferred from the observational data, the standard model (SBBN: \Nnu 
= 3) can be tested and nonstandard models (\Nnu $\neq 3$; $S \neq 1$) 
can be constrained.  As Figure~\ref{fig:hevsd234} reveals, the current 
data favor a slower than standard expansion rate (\Nnu $< 3$).  If both 
$\eta$ and \Deln are allowed to be free parameters, it is not surprising 
that the adopted primordial abundances of D and \4he can be accomodated 
(D largely fixes $\eta$, while \4he sets $S$).  In Figure 
\ref{fig:bbncontours} are shown the 1-, 2-, and 3-$\sigma$ BBN contours 
in the $\eta$ -- \Deln plane derived from the adopted values of \yd and 
\Yp.  Although the best-fit point is at \Deln $= -0.7$ ($S = 0.94$), it 
is clear that SBBN (\Nnu = 3) is not unacceptable. 

\section{BBN And The CBR}\label{sec:cbr}

\begin{figure}[t]  
\centering
 \epsfysize=3.5truein
  \epsfbox{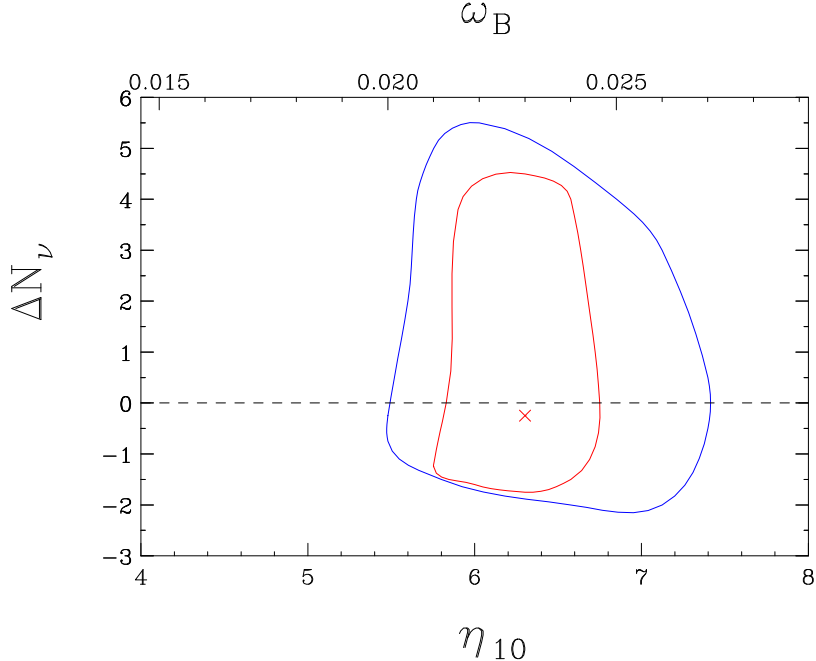}
\caption{The 1- and 2-$\sigma$ contours in the $\eta$ --  
\Deln plane from the CBR ({\it WMAP}) data.
\label{fig:wmapcontours}}   
\end{figure}

As for BBN, the CBR temperature anisotropy spectrum and polarization 
are also sensitive to the early-Universe (RD) expansion rate (see, 
\eg Barger \etal\cite{barger}, and references to related work therein). 
Increasing the baryon density increase the inertia of the baryon-photon
fluid, shifting the positions and the relative sizes of the odd and 
even acoutic peaks.  Changing the expansion rate (or, the relativistic
particle content) of the early Universe shifts the redshift of
matter-radiation equality so that there is a degeneracy between 
\Deln (or $S$) and the total matter density $w_{\rm M} \equiv 
\Omega_{\rm M}h^{2}$. It is the HST Key Project determination of 
the Hubble parameter that helps to break this degeneracy, permitting 
the WMAP data to set constraints on the post-BBN, early-Universe 
expansion rate.  Although the CBR temperature anisotropy spectrum 
is a less sensitive probe of the early-Universe expansion rate than 
is BBN, the WMAP data may be used to construct likelihood contours 
in the $\eta$ -- \Deln plane similar to those from BBN in Figure 
\ref{fig:bbncontours}. These CBR-derived contours are shown in Figure 
\ref{fig:wmapcontours}; note the very different \Deln scales and 
ranges in Figures \ref{fig:bbncontours} and \ref{fig:wmapcontours}.  
As is the case for BBN, the ``best'' fit value for the expansion rate 
is at $S < 1$ (\Deln $< 0$), but the CBR likelihood distribution of 
\Deln values is very shallow and the WMAP data is fully consistent 
with $S = 1$ (\Deln = 0).

\begin{figure}[t]  
\centering
 \epsfysize=3.5truein
  \epsfbox{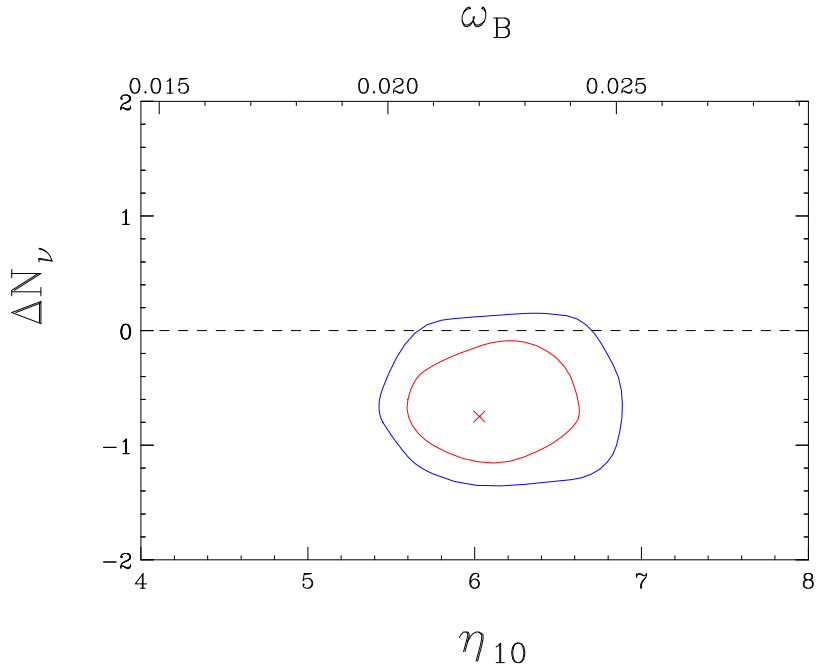}
\caption{The 1- and 2-$\sigma$ contours in the $\eta$ --  
\Deln plane for the joint BBN -- CBR ({\it WMAP}) fit. 
\label{fig:jointcontours}}   
\end{figure}

Comparing Figures \ref{fig:bbncontours} and \ref{fig:wmapcontours}, it 
is clear that there is excellent overlap between the $\eta$ -- \Deln 
confidence contours from BBN and those from the CBR (see Barger 
\etal\cite{barger}).  This nonstandard variant of SBBN ($S \neq 1$) is
also consistent with the CBR.  In Figure \ref{fig:jointcontours} (from 
Barger \etal\cite{barger}) are shown the confidence contours in the 
$\eta$ -- \Deln plane for a joint BBN -- CBR fit.  Again, while the 
best fit value for \Deln is negative (driven largely by the adopted 
value for \Yp), \Deln = 0 ($S = 1$) is quite acceptable. 

\section{Summary And Conclusions}\label{summary}

Adopting the standard models of cosmology and particle physics, SBBN 
predicts the primordial abundances of D, \3he, \4he, and \7li, which 
may be compared with the observational data.  Of the light nuclides, 
deuterium is the baryometer of choice, while \4he is an excellent 
chronometer.  The universal density of baryons inferred from SBBN 
and the adopted primordial D abundance: $\eta_{10}({\rm SBBN}) = 
6.10^{+0.67}_{-0.52}$, is in excellent agreement with the baryon 
density derived largely from CBR data~\cite{wmap}: $\eta_{10}({\rm 
CBR}) = 6.14 \pm 0.25$.  For this baryon density, the predicted 
primordial abundance of \3he is also in excellent agreement with 
the primordial value inferred from observations of an outer-Galaxy 
\hii region (Bania \etal \cite{gg}).  In contrast, the SBBN-predicted 
mass fraction of \4he for the concordant baryon density is Y$_{\rm 
P}^{\rm SBBN} = 0.248 \pm 0.001$, while that inferred from observations 
of recombination lines in metal-poor, extragalactic \hii regions 
\cite{osw} is lower: Y$_{\rm P}^{\rm OSW} = 0.238 \pm 0.005$.  
Since the uncertainties in the observationally inferred primordial 
value are likely dominated by systematics, this $\sim 2\sigma$ difference 
may not be cause for too much concern.  If it survives the accumulation 
of new data, this tension between D and \4he (or between the CBR-determined 
baryon density and \4he) can be relieved by a variation of the standard 
model in which the early, RD Universe expands more slowly than it should 
according to the standard model.  If {\it both} the baryon density and 
the expansion rate factor are allowed to be free parameters, BBN (D,
\3he, and \4he) and the CBR (WMAP) agree (at 95\% confidence) for $5.5 
\leq \eta_{10} \leq 6.8$ ($0.020 \leq \Omega_{\rm B}h^{2} \leq 0.025$) 
and $1.65 \leq$~N$_{\nu} \leq 3.03$.  More D and \4he data, along 
with a better understanding of systematic errors, will be crucial to 
deciding whether the source of this tension is inaccurate data (and/or 
data analysis) or is the first hint of new physics. 
 
The wealth of observational data accumulated in recent years have propelled 
the study of cosmology from youth to maturity.  In the current, data-rich 
era of cosmology, BBN continues to play an important role.  The spectacular 
agreement of the baryon density inferred from processes occurring at widely 
separated epochs confirms the general features of the standard models of 
cosmology and particle physics.  The tension with \4he provides a challenge, 
along with opportunities, to cosmology, to astrophysics, and to particle 
physics.  Whether the resolution of the current challenge is  observational 
or theoretical, the future is bright.

\section*{Acknowledgments}
I am grateful to all my collaborators, past and present, and I thank 
them for their contributions to the material reviewed here.  Many of 
the quantitative results (and figures) presented here are from recent 
collaborations or discusions with V. Barger, J. P. Kneller, D. Marfatia, 
K. A. Olive, R. J. Scherrer, and T. P. Walker.  My research is supported 
at OSU by the DOE through grant DE-FG02-91ER40690. This manuscript was 
prepared while I was visiting the Instituto Astr$\hat{\rm o}$nomico e 
Geof\' \i sico of the Universidade de S$\tilde{\rm }$ao Paulo; I thank 
them for their hospitality.

\end{document}